# Origin of the high DC transport critical current density for the MgB$_2$ superconductor

Kijoon H. P. Kim, W. N. Kang, Mun-Seog Kim, C. U. Jung, Hyeong-Jin Kim, Eun-Mi Choi, Min-Seok Park & Sung-Ik Lee

*National Creative Research Initiative Center for Superconductivity, Department of Physics, Pohang University of Science and Technology, Pohang 790-784, Korea*

**If the critical current density $J_c$ is very high for a superconductor, then estimating its value from transport measurements is not very easy. In such cases, the value of $J_c$, called $J_c^m$ for magnetic $J_c$, is usually obtained from the measured magnetic hysteresis loop measurements by using a proper critical state model such as Bean's model (ref. 1). However, for bulk polycrystalline high temperature superconductors, the values of $J_c^m$ are much higher than the values, $J_c^{tr}$, obtained from the transport measurements. This is due to the fact that the $J_c^{tr}$ is interrupted by weakly linked grain boundaries. However, for the recently discovered superconductor MgB$_2$ (ref. 2), the grain boundary effect is negligible and these two values seem to coincide. Moreover, $J_c$ increases drastically with decreasing the temperature. Consequently, the critical current densities for bulk wires can be very high, suggesting that numerous applications for the power transport. In this letter, we report the origin of the large current carrying capability of MgB$_2$ based on direct measurements of the current-voltage relation in high magnetic fields. A strong coupling between the grains may be one reason for the absence of the weak link effect. Another reason may be the fact that, instead of a weak pinning mechanism such as thermally activated flux hopping, strong pinning due to a vortex glass phase is found in this material. The vortex phase diagram obtained from transport measurements shows that, in H-T space, a wide region below the $H_{c2}$ line is covered by a vortex glass phase.**

Three-dimensional (3D) bulk polycrystalline samples (4.5 mm in diameter and 3.3 mm in height) were sintered at 950 °C under a pressure of 3 GPa. The transmission electron microscope images did not show any pores in the specimen. All grains were compactly connected, and no discernable empty spaces or impurities were found at the boundaries.[3] This is in strong contrast to the weakly connected morphology of the high temperature cuprate superconductors. In order to obtain a higher longitudinal voltage signal, we cut the sample into a bar shape 4 mm in length x 460 μm in width x 70 μm in thickness. The standard photolithography technique was adopted for fabricating the electrical contact pads. To obtain good ohmic contacts (< 1 Ω), we coated Au film on the contact pads after cleaning the sample surface with an Ar ion beam. Voltage noise, which is detrimental to precise measurements, was successfully reduced to a lower level.

The inset of Fig. 1 shows the superconducting resistive transition for $0 \leq H \leq 5$ T with a bias current density $J = 3.1$ A/cm$^2$. The zero-field onset $T_c$ is 38.5 K with a narrow transition width ~0.5 K, as judged from the 90% to 10% superconducting transition. Figure 1 shows the temperature dependence of the $J_c$ for $H = 0.5 - 5$ T. The values ($J_c^m$) for $J_c > 10^3$ A/cm$^2$ were estimated from the hysteresis in the magnetization measurements by using the Bean critical model[1] while those ($J_c^{tr}$) for $J_c < 10^3$ A/cm$^2$ were estimated from transport measurements using a 1 μV/mm criterion. Especially, the value of $J_c^{tr}$ at 21 K for $H = 5$ T was about $2.3 \times 10^2$ A/cm$^2$. The $J_c$'s obtained from both limits agreed with each other, which they do not do for weak-linked cuprate superconductors.[4,5] This implies that the bulk $J_c^{tr}$ could be very large, thus, large current transport through a power line becomes possible.

The question arises as to why $J_c^{tr}$ increases rapidly by several orders of magnitude with decreasing temperature. This answer should be found in the response of the vortex to the current. For this purpose, we study the *I-V* characteristics for fields in the range $0 \leq H \leq 5$ T. Figure 2(a) shows *I-V* curves for temperatures from 26.8 K to 33.0 K for H = 3 T, which were very similar to those of the cuprate superconductor near the vortex glass transition point, $T_g$. An analysis based on the vortex-glass to vortex-liquid transition theory[6,7] was performed. According to the vortex glass theory, *I-V* curves collapse into a scaling function near the vortex-glass phase transition after transforming *V* and *I* into two variables $V_{sc}=V/I|T-T_g|^{\nu(z-1)}$ and $I_{sc}=I/T|T-T_g|^{2\nu}$, respectively. The scaling functions were obtained as shown in Fig. 2(*b*) with critical exponents $\nu = 1.5$ and $z = 2.3$, which were in good agreement with the theoretical prediction for a 3D system. The existence of this scaling behavior in a bulk 3D polycrystalline sample is notable because normally this behavior is screened by the grain boundary effect, as in YBa$_2$Cu$_3$O$_x$. This even supports the idea[8,9] that the grain boundary does not function as a weak linker in MgB$_2$. In a vortex glass state, the pinning potential becomes extremely strong at low temperatures compared to that in the thermally activated region, which is the reason $J_c$ increases drastically with decreasing temperature. The weak pinning behavior, which appears in thermally activated vortex-hopping phenomena, is not observed in a vortex solid state.

Figure 3 shows the phase diagram in the *H-T* plane. The diagram is based on the vortex glass line $H_g(T)$, obtained from the analysis using the vortex-glass theory. The upper critical fields, $H_{c2}(T)$, were estimated from the *R-T* curves when the resistance drops to 90% of the normal-state resistance. The magnetic field dependence of $T_g$ was well described by $H_g \propto (1-T_g/T_c)^n$ with $n = 1.38$. This value is almost the same as those reported in high-T$_c$ cuprate superconductors, such as YBa$_2$Cu$_3$O$_x$ single crystals,[10] and is consistent with the theoretical value, $n = 4/3$.[6]

**Acknowledgments**
This work is supported by the Ministry of Science and Technology of Korea through the Creative Research Initiative Program.

Correspondence and requests for materials should be addressed to Sung-Ik Lee
(e-mail: silee@postech.ac.kr)

**Figure captions**

Fig. 1. Temperature dependence of the critical current density of a 3D polycrystalline $MgB_2$ sample for H = 0.5, 1.0, 2.0, 3.0, and 5.0 T. The values for $J_c > 10^3$ A/cm$^2$ were estimated from the hysteresis in the magnetization measurements by using the Bean critical model[1] while those for $J_c < 10^3$ A/cm$^2$ were estimated from transport measurements using a 1-$\mu$V/mm criterion. The inset shows the resistive superconducting transition for H = 0.0, 0.1, 0.2, 0.5, 1.0, 2.0, 3.0, 4.0, and 5.0 T with a bias current density $J$ = 3.1 A/cm$^2$. The zero field onset $T_c$ is 38.5 K with a narrow transition width of ~0.5 K, as judged from the 90% to 10% drop of the resistive transition.

Fig. 2. (a) Current-voltage characteristics at $T$ = 26.8, 27.2, 27.6, 28.0, 28.4, 28.8, 29.2, 29.6, and 33.0 K for $H$ = 3 T. The dotted line represents the *I-V* curve at the vortex-glass transition temperature, $T_g$. (b) Vortex-glass scaling behavior. The *I-V* curves collapse into a scaling function near the vortex-glass phase transition after transforming *V* and *I* into two variables $V_{sc}=V/I|T-T_g|^{\nu(z-1)}$ and $I_{sc}=I/T|T-T_g|^{2\nu}$, respectively. The scaling exponents, $\nu$ = 1.5 and $z$ = 2.3, are in good agreement with the theoretical predictions for a 3D system.[6,7]

Fig. 3. Phase diagram for $MgB_2$ based on a vortex-glass to vortex-liquid transition. $H_g(T)$ was obtained using scaling procedures, and $H_{c2}(T)$ were estimated from the *R-T* curves when the resistance drops to 90% of the normal-state resistance.

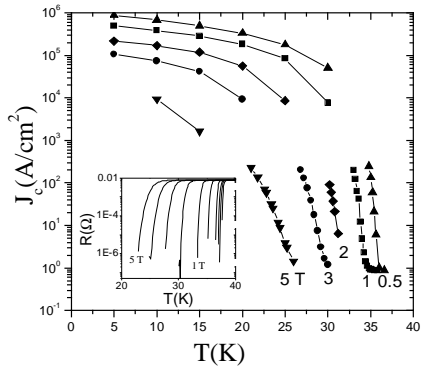

Figure 1.  Kijoon H. P. Kim *et al.*

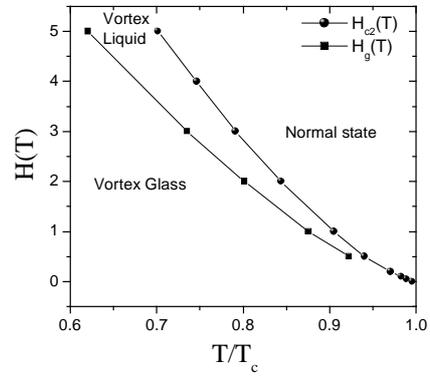

Figure 3.  Kijoon H. P. Kim *et al.*

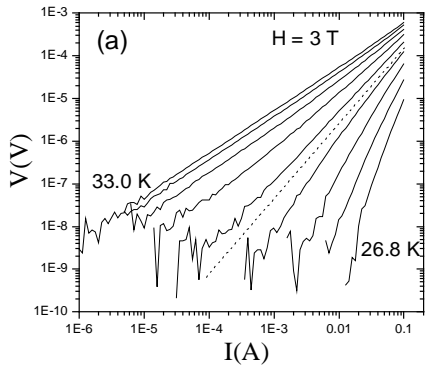

Figure 2(a).  Kijoon H. P. Kim *et al.*

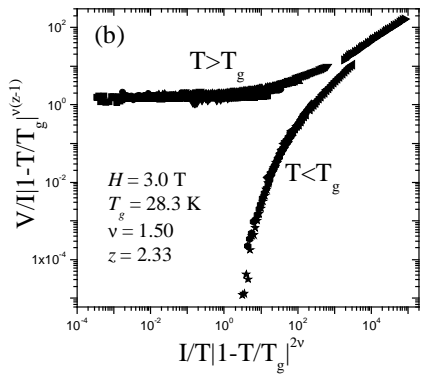

Figure 2(b).  Kijoon H. P. Kim *et al.*

3